\RequirePackage[mathlines,displaymath]{lineno}
\documentclass[twocolumn,prd,superscriptaddress,altaffilletter,nofootinbib]{revtex4-1}

\usepackage[mathlines,displaymath]{lineno}
\usepackage{graphicx}   

\usepackage{subcaption}  
\usepackage{hyperref} 
\usepackage{physics}
\usepackage{color}
\usepackage{amssymb}
\usepackage[utf8]{inputenc}
\DeclareUnicodeCharacter{00B4}{\textvisiblespace}

\begin{document}

\title{Subtracting compact binary foreground sources to reveal primordial gravitational-wave backgrounds}
\author{Surabhi Sachdev}
\email{szs1416@psu.edu}
\affiliation{Institute for Gravitation and the Cosmos, Physics Department, Pennsylvania State University, University Park, PA, 16802, USA}

\author{Tania Regimbau}
\email{tania.regimbau@lapp.in2p3.fr}
\affiliation{LAPP, Universit'{e} Grenoble Alpes, USMB, CNRS/IN2P3, F-74000 Annecy, France}

\author{B. S. Sathyaprakash}
\email{bss25@psu.edu}
\affiliation{Institute for Gravitation and the Cosmos, Physics Department, Pennsylvania State University, University Park, PA, 16802, USA}
\affiliation{Department of Astronomy \& Astrophysics, Pennsylvania State University, University Park, PA, 16802, USA}
\affiliation{School of Physics and Astronomy, Cardiff University, Cardiff, UK, CF24 3AA}

\begin{abstract}
Detection of primordial gravitational-wave backgrounds generated during the
early universe phase transitions is a key science goal for future ground-based
detectors. The rate of compact binary mergers is so large that their cosmological 
population produces a confusion background that could masquerade the detection of 
potential primordial stochastic backgrounds. In this paper we study the ability 
of current and future detectors to resolve the confusion background to reveal 
interesting primordial backgrounds. The current detector network of LIGO and 
Virgo and the upcoming KAGRA and LIGO-India will not be able to resolve 
the cosmological compact binary source population and its sensitivity to stochastic
background will be limited by the confusion background of these sources. We find that 
a network of three (and five) third generation (3G) detectors of Cosmic Explorer and Einstein
Telescope will resolve the confusion background produced by binary black holes
leaving only about 0.013\% (respectively, 0.00075\%) unresolved; in contrast,
as many as 25\% (respectively, 7.7\%) of binary neutron star sources remain
unresolved.  Consequently, the binary black hole population will likely not
limit observation of primordial backgrounds but the binary neutron star
population will limit the sensitivity of 3G detectors to $\Omega_{\rm GW} \sim
10^{-11}$ at 10 Hz (respectively, $\Omega_{\rm GW} \sim 3\times 10^{-12}$). 
\end{abstract}

\maketitle

\date{\today}

\section{Introduction} \label{sec:introduction}
With the continued detections of gravitational waves
from binary black hole mergers~\cite{Abbott2016a, Abbott2016b, Abbott2017,
Abbott2017b, Abbott2017a, LIGOScientific:2018mvr} and binary
neutron star inspirals~\cite{PhysRevLett.119.161101, Abbott:2020uma}, the LIGO
Scientific and Virgo Collaborations have kept up to their promise of taking us
into an era of gravitational-wave astronomy. In addition to these loud and
nearby sources that are seen as isolated transient events, there is a
population of weak, unresolved sources at higher
redshifts~\cite{apj.739.86.11,prd.84.084004.11,prd.84.124037.11,prd.85.104024.12,mnras.431.882.13}.
The superposition of these sources is expected to be the main contributor to
the astrophysical stochastic background which may be detectable in the next few
years as the Advanced LIGO~\cite{Aasi2015} and Virgo
detectors~\cite{TheVirgoCollaboration2015} reach their design sensitivity and
accumulate more data~\cite{PhysRevLett.116.131102, PhysRevLett.120.091101}.
Assuming the most probable rate for compact binary mergers at the time ($103^{+110}_{-63}
\mathrm{Gpc}^{-3} \mathrm{yr}^{-1}$~\cite{Abbott2017} for BBH and
$1540_{-1220}^{+3200}
\textrm{Gpc}^{-3}\textrm{yr}^{-1}$~\cite{PhysRevLett.119.161101} for BNS),  it
has been shown that the total background may be detectable with a
signal-to-noise-ratio  of  3  after 40  months  of  total  observation  time,
based  on  the  expected  timeline for Advanced LIGO and Virgo to reach their
design sensitivity~\cite{PhysRevLett.120.091101}. The astrophysical background
potentially contains a wealth of information about the history and evolution of
a population of point sources, but it is a confusion noise background that
obscures the observation of the primordial gravitational-wave background (PGWB)
produced in the very early stages of the Universe. Proposed theoretical
cosmological models include the amplification of vacuum fluctuations during
inflation~\cite{spjetp.40.409.75, jetpl.30.682.79, prd.48.3513.93}, pre
Big-Bang models~\cite{app.1.317.93,prd.55.3330.97,prd.82.083518.10}, cosmic
(super)
strings~\cite{prd.71.063510.05,prl.98.111101.07,prd.81.104028.10,prd.85.066001.12}
, or phase transitions~\cite{prd.77.124015.08,prd.79.083519.09,jcap.12.024.09}.
For a comprehensive discussion of cosmological gravitational-wave backgrounds, 
we refer the reader to reviews by Maggiore and \cite{pr.331.283.00} and 
Bin\'etruy et al. \cite{jcap.6.27.12}.

Detection of the primordial gravitational-wave background would create a unique
window on the very first instants of the Universe, up to the limits of the
Planck era, and on the physical laws that apply at the highest energy scales.
Needless to say that such a detection would have a profound  impact  on  our
understanding  of  the  evolution  of the  Universe.

In addition to the astrophysical background from unresolved compact binary
mergers, a contribution is expected to result from the superposition of several
other unresolved sources~\cite{raap.11.369.11}, such as cosmic (super)
strings~\cite{prl.98.111101.07}, core collapse supernovae to neutron stars or
black
holes~\cite{prd.72.084001.05,prd.73.104024.06,mnras.398.293.09,mnrasl.409.L132.10},
rotating neutron stars~\cite{aap.376.381.01,prd.86.104007.12} including
magnetars~\cite{aap.447.1.06,mnras.410.2123.10,mnras.411.2549.11,prd.87.042002.13},
phase transitions~\cite{grg.41.1389.09}, or initial instabilities in young
neutron stars~\cite{mnras.303.258.99,mnras.351.1237.04,apj.729.59.11}.
 
The current detector network of LIGO and Virgo and the upcoming KAGRA and
LIGO-India will not be able to resolve the cosmological compact binary source
population and its sensitivity to stochastic background will be limited by the
confusion background of these sources~\cite{PhysRevLett.118.151105}. With the
increased sensitivity of the third generation gravitational-wave detectors,
such as the Einstein Telescope (ET)~\cite{0264-9381-27-19-194002} and the
Cosmic Explorer (CE)~\cite{0264-9381-34-4-044001}, it will be possible to
detect and resolve almost all of the binary black hole mergers, even the ones
at high redshifts.  In this work, we explore the possibility of probing the
cosmological gravitational-wave background with the third generation detectors,
after removing the astrophysical background from compact binary mergers from
the data. This work is an extension to~\cite{PhysRevLett.118.151105}, where the
authors have shown the level at which we can expect amplitude of background
from unresolved, subthreshold signals from compact binary coalescences (CBC)
using different detector networks.  We extend the previous study to also
provide an estimate of errors we introduce while subtracting the signals
above threshold for the most optimistic network of detectors considered
by~\cite{PhysRevLett.118.151105}. The idea of subtracting foreground signals
to extract stochastic backgrounds was already explored \cite{Harms:2008xv} 
in the context of the the Big Bang Observer \cite{Crowder:2005nr}, including a 
noise projection method that could reduce errors due to imperfect 
subtraction \cite{Cutler:2005qq}.

Data from gravitational-wave detectors are dominated by
environmental and instrumental backgrounds. Consequently, it is not possible to
identify even deterministic signals without sophisticated data processing such
as matched filtering \cite{Sathyaprakash:1991mt}.  Stochastic backgrounds cannot
be reliably detected in a single detector---they are found by cross-correlating
the data from a pair of detectors. Indeed, the stochastic background present in
one of the detectors acts as a matched filter for the data in the other detector
\cite{Schutz:1987ru,1999PhRvD..59j2001A,prd.59.102001.99}. 
Unfortunately, this means that any common
noise in a pair of detectors could masquarade as stochastic background
\cite{Aasi:2014jkh}.  If detectors are geographically well separated then the
risk of common noise of terrestrial origin is greatly reduced. Additionally,
certain backgrounds of terrestrial origin could be measured and subtracted 
\cite{2018PhRvD..97j2007C}.  Even in the
absence of any terrestrial background, a pair of detectors would see the same
astrophysical background, which would show up as correlated `noise' although
detectors might be geographically well separated. As a result, the only possible
way to improve the sensitivity of a detector network to primordial backgrounds
is to subtract foreground astrophysical signals.

The rest of the paper is organized as follows.
In Sec.~\ref{sec:methods}, we describe the basic method that we use to calculate
the gravitational-wave spectrum from the error introduced by imperfect
subtraction of CBC signals. In Sec.~\ref{sec:PE}, we describe the framework used to
estimate the deviations of the estimated parameters of the CBC sources from
their true values. We discuss the simulation of a population
of binaries in Sec.~\ref{sec:popsynth}, discuss the result of the imperfect subtraction
of such signals in Sec.~\ref{sec:simulations}, and we discuss our results in Sec.~\ref{sec:discussion}.

\section{Method}\label{sec:methods}

The energy-density spectrum in gravitational waves is described by the
dimensionless quantity~\cite{prd.59.102001.99},
\begin{equation}
\Omega_\text{GW}(f) =
\frac{f}{\rho_c}\frac{d\rho_{\text{GW}}}{df},
\end{equation}
where $d\rho_{\text{GW}}$ is the energy density in the frequency interval $f$
to $f+df$, $\rho_c = 3H_0^2c^2/8\pi G$ is the closure energy density, and $H_0$
is the Hubble constant equal to $67.8\pm0.9$ km/c/Mpc~\cite{refId0}.

The gravitational-wave energy spectrum density can be written as a sum of
contribution from the astrophysical and cosmological energy densities,
\begin{equation}
\Omega_\text{GW} = \Omega_\text{astro}+\Omega_\text{cosmo}.
\end{equation}
Taking the contribution of the compact binary coalescences out of the
astrophysical background, and writing it explicitly, we have,
\begin{align}
\Omega_\text{GW} = \Omega_\text{astro, r}+\Omega_\text{cosmo}+\Omega_\text{cbc}.
\end{align}
Here $\Omega_\text{astro, r}$ is the remaining astrophysical background after
taking out the contribution from the CBC sources.

When estimating the parameters of a binary source, by using Monte Carlo
methods, or nested sampling, we invariably end up with parameters that deviate
from the true values because of the noise in the detector. Therefore when we
subtract the recovered CBC signals from the data, we introduce an additional
background due to the error in subtraction, $\Omega_\text{error}$.

\begin{align}
\Omega_\text{GW}& = \Omega_\text{cbc, rec}+\Omega_\text{error}\\ \nonumber 
& + \Omega_\text{cbc, unres} + \Omega_\text{cosmo} +
\Omega_\text{astro, r},
\end{align}
where $\Omega_\text{cbc, rec}$ is the background from the recovered CBC sources
that we can subtract from our data, $\Omega_\text{error}$ is the background
because of the error introduced from such a subtraction, $\Omega_\text{cbc,
unres}$ is the background from the unresolved CBC sources which are not
detected as foreground events. Let us assume that we have an experiment where
we have detected a list of CBC sources and subtracted them from the data. Now
we are left with the gravitational-wave backgrounds, $\Omega_\text{error}$,
$\Omega_\text{cbc, unres}$, on top of the cosmological and astrophysical (from
sources other than the CBCs) backgrounds. We want to answer the question of
whether the cosmological or astrophysical backgrounds from sources other than
CBCs can stand above the residual background after removal of the CBC sources.
That is,
\begin{align}
\Omega_\text{error}+\Omega_\text{cbc, unres}\stackrel{\text{?}}{\lessgtr}\Omega_\text{cosmo}\stackrel{\text{?}}{\lessgtr}\Omega_\text{astro, r}.
\end{align}
In order for us to be able to detect the gravitational-wave background from
cosmological sources or that from different astrophysical sources, we would
need $\Omega_\text{residual} = \Omega_\text{error}+\Omega_\text{cbc, unres}$ to
lie below these.

The gravitational-wave energy density from a population of compact binary
sources is given by~\cite{PhysRevLett.118.151105},
\begin{align}
\Omega_\text{cbc} & = \frac{1}{\rho_c c}fF(f),
\end{align} 
where $F(f)$ is the total flux, sum of individual contributions
\begin{align}
F(f) & = T^{-1} \frac{\pi c^3}{2G}f^2\sum_{k=1}^N(\tilde{h}^2_{+, k}(f)
+\tilde{h}^2_{\times, k}(f)),
\end{align} 
where $N$ is the number of sources in the Monte Carlo sample, and $T^{-1}$
assures that flux has the correct dimension, $T$ being the total time of the
data sample. $\tilde{h}_{+, k}(f)$ and $\tilde{h}_{\cross, k}(f)$ are the Fourier
domain waveforms for the two polarizations, and the index $k$ runs over all the
sources.  We calculate $\Omega_\text{ error}$ as,
\begin{equation}
\Omega_\text{error}  = \frac{1}{\rho_c c}fF_\text{error}(f), \\
\end{equation}
where,
\begin{align}
F_\text{error}(f)& = T^{-1} \frac{\pi
c^3}{2G}f^2\sum_{k=1}^N((\tilde{h}^\text{true}_{+,
k}(f)-\tilde{h}^\text{recovered}_{+, k}(f))^2\\ \nonumber
&+(\tilde{h}^\text{true}_{\times, k}(f)-\tilde{h}^\text{recovered}_{\times,
k}(f))^2).
\end{align}

To get an estimate of $\Omega_\text{error}$, we need to estimate the
quantities, $\tilde{h}^\text{recovered}_{+, k}(f)$ and
$\tilde{h}^\text{recovered}_{\times, k}(f)$.

\section{Estimating the deviation from true value of the measured source parameters}
\label{sec:PE} 
Ideally we want the full Bayesian posteriors to estimate
the deviation from the true value of parameters. However, at present it is
unfeasible to compute the full posterior probability
distribution functions of all $15$ binary parameters for the hundreds of
thousands of sources that we simulate up to a redshit of $10$ in the following
section. The Fisher matrix provides a computationally inexpensive method to estimate
the errors in the case when the posteriors are Gaussian, which is, unfortunately,
not true in general. Nevertheless, for the purpose of building a proof-of-principle
concept the Fisher matrix method is adequate and the only practical approach to obtain the 
magnitude of errors in the estimation of parameters. To this end, 
we follow the framework described in~\cite{arun2005} and calculate the errors
in estmating the parameters of the compact binary system using the Fisher
matrix method.\\

According to the post-Newtonian expansion
formalism~\cite{buonanno2009comparison}, the gravitational-wave strain from a
compact binary coalescence in frequency domain is given by
\begin{align}
\tilde{h}(f) & = \mathcal{A}f^{-7/6}e^{i\Psi(f)}, 	\label{0pn}
\end{align}
where $A$ is the amplitude of the waveform, and $\Psi(f)$ is the
phase given by
\begin{align}
\Psi(f) & = 2 \pi ft_{c} - \phi_{c} - \frac{\pi}{4} +
\frac{3}{128\eta\nu^{5}}\sum_{k=0}^{N}\alpha_{k}\nu^{k}.
\end{align}
Here $t_c$ is the time of coalescence, $\phi_c$ is the coalescence phase, $\nu
= (\pi Mf)^{1/3}$, $M$ is the total mass ($M = m_1+m_2$), $\eta$ is the symmetric
mass ratio ($\eta = m_1m_2/M^2$) of the system, and the $\alpha_{k}$ terms are
known as the post-Newtonian (PN) coefficients. In this work, we restrict
ourselves to 0-PN approximation (or the Newtonian approximation, $k=0$),
which will be justified below.  For the Fisher matrix study, we choose 
a set of independent parameters $\vec{\theta}$ for describing the 
gravitational waveform,
\begin{equation}
\vec{\theta} = (f_{0}t_{c}, \phi_{c}, \ln\mathcal{M}),
\end{equation}
where $f_0$ is a reference frequency needed to keep the parameters for the Fisher
matrix dimensionless. $\mathcal{M}$ is the dimensionless chirp mass, and is
defined as $\mathcal{M} = \eta^{3/5}M/M_\odot$.

Writing the phase of the waveform in terms of these parameters, we have,
\begin{equation}
\Psi(f) = 2 \pi \frac{f}{f_{0}}(f_{0}t_{c}) - \phi_{c} -
\frac{\pi}{4} + \frac{3}{128}\left (\pi \mathcal{M}f \right )^{-5/3},  \\ \label{waveform}
\end{equation}
or equivalently,
\begin{equation}
\Psi(f; \vec\theta) = 2 \pi \frac{f}{f_{0}}\theta_1 - \theta_2
- \frac{\pi}{4} + \frac{3}{128}\left (\frac{\pi e^{\theta_3}fGM_\odot}{c^3} \right)^{-5/3}. \label{phasewfs}
\end{equation}
In going from Eq. (\ref{waveform}) to Eq. (\ref{phasewfs}), we have truncated
the expansion at $\alpha_0$ term, plugged in the value $\alpha_0 = 1$, and we
have introduced the Newton's constant $G,$ the speed of light $c,$ and solar
mass $M_{\odot}$, explicitly to keep all quantities in the Eq. (\ref{waveform})
dimensionless, and defined masses in solar mass units.

The Fisher matrix elements are given by,
\begin{equation} \label{gamma-eq}
\Gamma_{ij} = 2\int_{f_L}^{f_H}
\frac{\tilde{h}_{\theta_i}^{*}(f; \vec\theta)\tilde{h}_{\theta_j}(f; \vec\theta) +
\tilde{h}_{\theta_i}(f; \vec\theta)\tilde{h}_{\theta_j}^{*}(f; \vec\theta)}{S_{n}(f)}\mathrm{d}f,
\end{equation}
where
\begin{equation}
\tilde{h}_{\theta_i}(f; \vec\theta) =
\frac{\partial\tilde{h}(f; \vec\theta)}{\partial{\theta_i}}
\end{equation}
are the partial derivatives of the waveform with respect to $\theta_i$, the
parameters of the waveforms, and $S_n(f)$ is the single-sided power spectral
density of the detector.  The partial derivatives of the waveform can be
calculated analytically:
\begin{equation}
\tilde{h}_{\theta_1}(f; \vec\theta) = \frac{2\pi
fA}{f_0}f^{-7/6}e^{i(\Psi(f; \vec\theta)+\pi/2)},
\end{equation}
\begin{equation}
\tilde{h}_{\theta_2}(f; \vec\theta) = Af^{-7/6}e^{i(\Psi(f; \vec\theta)-\pi/2)},
\end{equation}
and,
\begin{equation}
\tilde{h}_{\theta_3}(f; \vec\theta) =
Af^{-7/6}e^{i(\Psi(f; \vec\theta)-\pi/2)}\frac{5}{128}\left (
\frac{\pi e^{\theta_3}fG}{c^3} \right )^{-5/3}.
\end{equation}
The Fisher matrix is then calculated by performing the integration in Eq.
(\ref{gamma-eq}) numerically.  For a network of detectors, the Fisher matrix is
the sum of Fisher matrices for individual detectors, 
\begin{align}
\Gamma^\text{net}_{ij} = \sum_\text{det}\Gamma^\text{det}_{ij}.  
\end{align}
The variance-covariance matrix, or simply the covariance matrix, defined as
the inverse of the Fisher information matrix, is given by
\begin{equation}
\Sigma_{ij} = (\Gamma^{-1})_{ij}.
\end{equation}

Once we have the covariance matrix, we use a multivariate normal random number
generator to generate observed values of the parameters, $\mathbf{P_O}$, based on
the multivariate dirstribution with the mean equal to the true value of the
parameters, $\mathbf{P_T}$ and covariance matrix as $\Sigma$.  The error in
parameter estimation is then given by
\begin{align}
\mathbf{R}  = [\Delta
\theta_1, \Delta \theta_2, \Delta \theta_3] & = \mathbf{P_O}-\mathbf{P_T},
\end{align}
where
\begin{align}
\Delta t_c = \frac{\Delta \theta_1}{f_0},  \quad
\Delta \phi_{c} = \Delta \theta_2, \quad
\Delta \mathcal{M} = \mathcal{M} \Delta \theta_3.  
\end{align}

\section{Population synthesis for multiple detectors}\label{sec:popsynth}
We simulate a population of binary black hole and binary neutron star systems
up to a redshift of 10, and then calculate an estimate of
$\Omega_\text{cbc, rec}$ and $\Omega_\text{error}$ as outlined in Sec.~\ref{sec:methods} 
and Sec.~\ref{sec:PE}. The list
of compact binaries (neutron star binaries or black hole binaries) is generated
following a Monte Carlo procedure described in~\cite{2012PhRvD..86l2001R,
2014PhRvD..89h4046R, 2015PhRvD..92f3002M, PhysRevLett.118.151105}, and using
the fiducial model of~\cite{PhysRevLett.120.091101} for the distribution of the
parameters (masses, redshift, position on the sky, polarization and inclination
angle of the binary).  In particular, we assume a redshift distribution which
is derived from the star formation rate (SFR) of~\cite{vangioni2015} and accounts for
a delay between the formation of the progenitors and the merger. We further
consider the median rates estimated from the first LIGO observation run.

\begin{enumerate}
\item For BBHs, the intrinsic masses $m_1, m_2$ (in the source frame) are
selected from the power-law distribution (Saltpeter initial mass
function~\cite{1955ApJ...121..161S}) considered in~\cite{PhysRevX.6.041015,
Abbott2017} of the primary (i.e., the larger mass) companion
$p(m_1) \propto m_1^{-2.35}$ and from a uniform distribution of the secondary
companion. In addition, we require that the component masses take values in the
range 5--50 $M_\odot$.

For BNSs, the intrinsic masses $m_1, m_2$ (in the source frame) are both drawn
from a Gaussian  distribution centred around 1.33 $M_\odot$ with a standard deviation of 0.09 $M_\odot$.

\item The redshift $z$ is drawn from a probability distribution $p(z)$ given
 by
\begin{equation}
p(z) = \frac{R_z(z)}{\int_0^{10} R_z(z)dz},
\end{equation}
obtained by normalizing the merger rate of binaries in the observer frame,
$R_z(z)$ per interval of redshift, over the range $z \in [0, 10]$. We choose to
cut off the redshift integral at $z_\mathrm{max} = 10$, since redshifts larger than 5
contribute little to the background~\cite{PhysRevLett.120.091101}.
The merger rate in the observer frame is\footnote{There was an error in Eq. 2 in ~\cite{PhysRevLett.118.151105}, we have corrected it here in Eq.~\ref{correct_rate_per_redshift}.}
\begin{equation}\label{correct_rate_per_redshift}
R_z(z) = \frac{R_m(z)}{1+z}\frac{dV}{dz}(z),
\end{equation}
where $dV/dz$ is the comoving volume element and $R_m(z)$ is the rate per
comoving volume in the source frame, given by
\begin{equation}\label{mergerRateSource}
R_m(z) = \int_{t_\mathrm{min}}^{t_\mathrm{max}}\int\displaylimits_{z_f=z(t_m-t_d)}R_f(z_f)p(t_d) dz_f dt_d,
\end{equation}
where $R_f(z_f)$ is the binary formation rate as a function of the redshift at
formation time, $z_f = z(t_f)$ is the source redshift at formation, $p(t_d)$ is
the distribution of the time delay $t_d$ between the formation and merger of
the binary, $z=z(t_m)$ is the source redshift at merger. The integration in Eq.~\ref{mergerRateSource} over $z_f$ is performed for all the redshifts corresponding to $t_f$ such that $t_m=t_f+t_d$. 

We consider a time delay distribution $p(t_d) \propto 1/t_d$, for
$t_\textrm{min} < t_d < t_\textrm{max}$. For BNS, we set $t_\mathrm{min} = 20
\mathrm{Myr}$~\cite{prd.92.063002.15, PhysRevLett.120.091101}, whereas for BBH,
we set $t_\mathrm{min} = 50 \mathrm{Myr}$~\cite{PhysRevLett.116.131102,
PhysRevLett.120.091101, 0004-637X-779-1-72}. The maximum time delay,
$t_\mathrm{max}$ is set to the Hubble time~\cite{apj.572.407.02,
1475-7516-2004-06-007, apj.648.1110.06, 0004-637X-779-1-72,
0004-637X-664-2-1000, NAKAR2007166, 0004-637X-675-1-566, 0004-637X-759-1-52,
0004-637X-779-1-72}.

We assume that the binary formation rate $R_f(z_f)$ scales with the SFR. We follow
the the cosmic star formation model of ~\cite{vangioni2015} which uses the
Springer-Hernquist functional form~\cite{doi:10.1046/j.1365-8711.2003.06499.x}
\begin{equation}
R_f(z) = \nu \frac{a e^{b(z - z_m)}}{a -b + b e^{(a(z-z_m)}},
\end{equation}
to fit to the GRB-based high-redshift SFR data of~\cite{Kistler:2013jza} but normalized based
on the procedure described in \cite{ 2041-8205-773-2-L22, 0004-637X-799-1-32}. 
This fit results in $\nu = 0.146 M_\odot / \textrm{yr} / \textrm{Mpc}^3, z_m = 1.72, a = 2.80, 
\textrm{and } b =2.46$~\cite{vangioni2015}. The value of
$R_m(z=0)$ is chosen as the local merger rate estimate from the LIGO-Virgo
observations.  For  the rate of BBH mergers, we use the most recent published
result associated  with  the  power-law  mass  distribution $56^{+44}_{-27}
\mathrm{Gpc}^{-3} \mathrm{yr}^{-1}$~\cite{LIGOScientific:2018mvr}. For the BNS
case, we set $R_m(z=0)$ to $920_{-790}^{+2220}
\textrm{Gpc}^{-3}\textrm{yr}^{-1}$ also from~\cite{LIGOScientific:2018mvr}.
Massive black holes are formed preferentially in low-metallicity
environments~\cite{TheLIGOScientific:2016htt, PhysRevLett.116.131102}. For systems
where at least one black hole has a mass larger than $30 M_\odot$, we re-weight
the star formation rate $R_f(z)$ by the fraction of stars with metallicities
less than half the solar metallicity~\cite{PhysRevLett.120.091101}.
Following~\cite{PhysRevLett.116.131102, PhysRevLett.120.091101}, we use the mean metallicity-redshift relation of
~\cite{doi:10.1146/annurev-astro-081811-125615}, and scale it upwards by a factor of 3 to account for local observations~\cite{vangioni2015, Belczynski2016}. 
\item The location on the sky, the cosine of the inclination angle, the polarization, and
the coalescence phase are drawn from uniform distributions.
\end{enumerate}

\subsection{Detector Network}
We consider two networks of third generation detectors: one with three total
detectors, out of which two have the sensitivity of CE located at LIGO Hanford
and LIGO Livingston locations and one with the sensitivity of ET located at the
location of Virgo; and a five-detector network with one detector with the sensitivity of ET at the location
of Virgo, and detectors with CE sensitivity at locations of LIGO Hanford, LIGO
Livingston, LIGO India, and KAGRA. We choose these configurations for the
detector-networks because it was shown in~\cite{PhysRevLett.118.151105}, that
the astrophysical ``confusion" background from unresolved BBH sources is
decreased by orders of magnitude, reaching $\Omega_\text{GW}(10 \text{Hz}) = 10^{-14} - 10^{-13}$ and $\Omega_\text{GW}(10 \text{Hz}) = 10^{-16} - 10^{-14}$ respectively.

\begin{figure*}
\begin{subfigure}[t]{0.49\textwidth}
\includegraphics[width=1.0\textwidth]{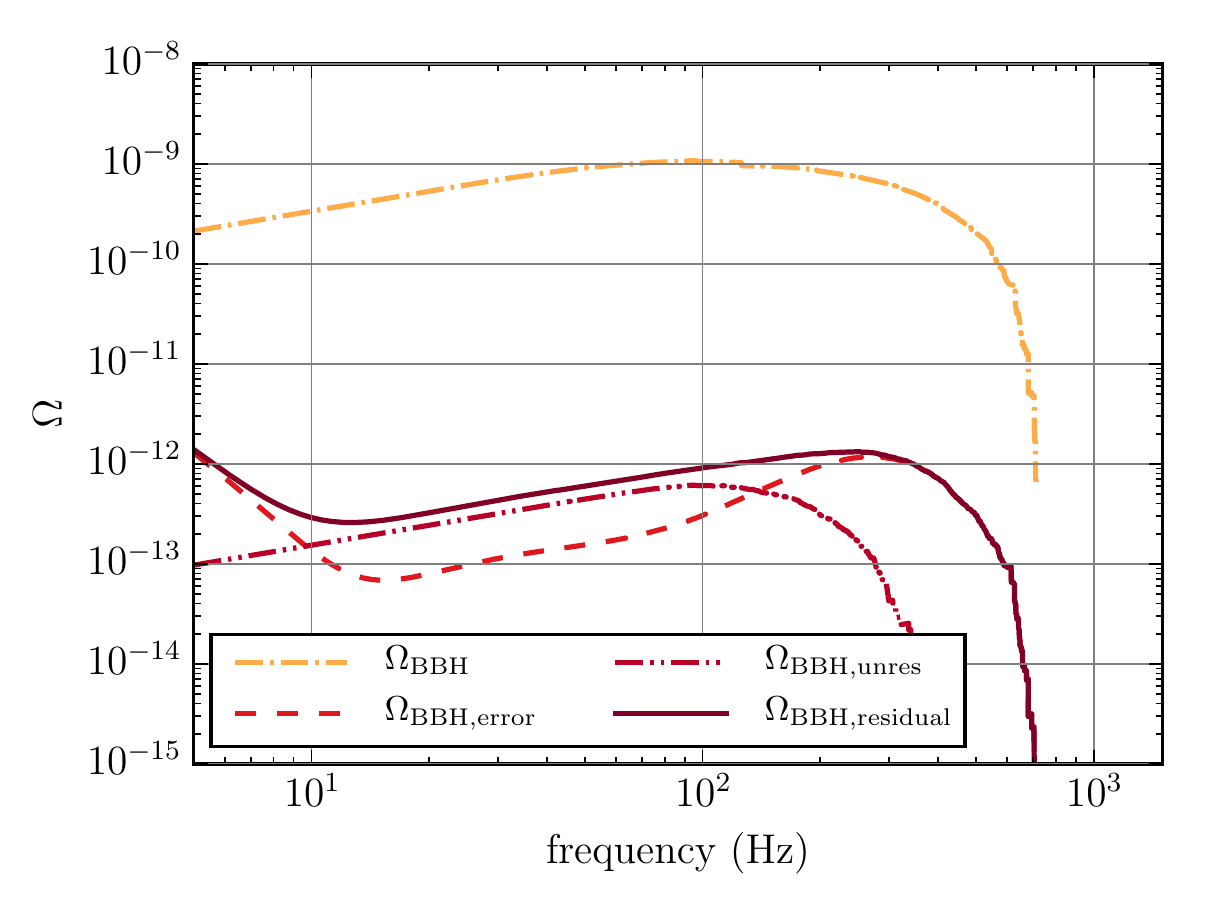} \caption{BBH, HLV network}
\end{subfigure}%
\hfill
\begin{subfigure}[t]{0.49\textwidth}
\includegraphics[width=1.0\textwidth]{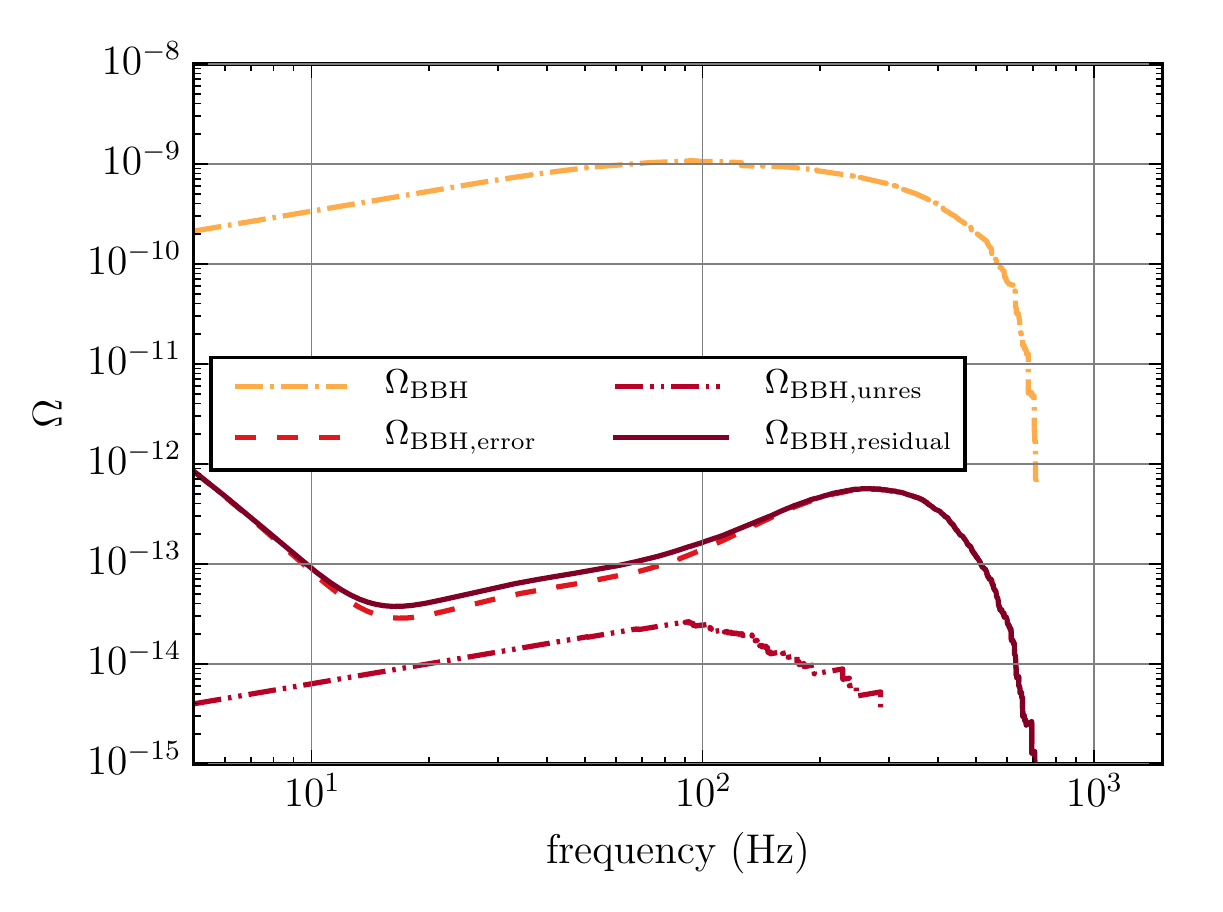} \caption{BBH, HLVIK network}
\end{subfigure}
\begin{subfigure}[t]{0.49\textwidth}
\includegraphics[width=1.0\textwidth]{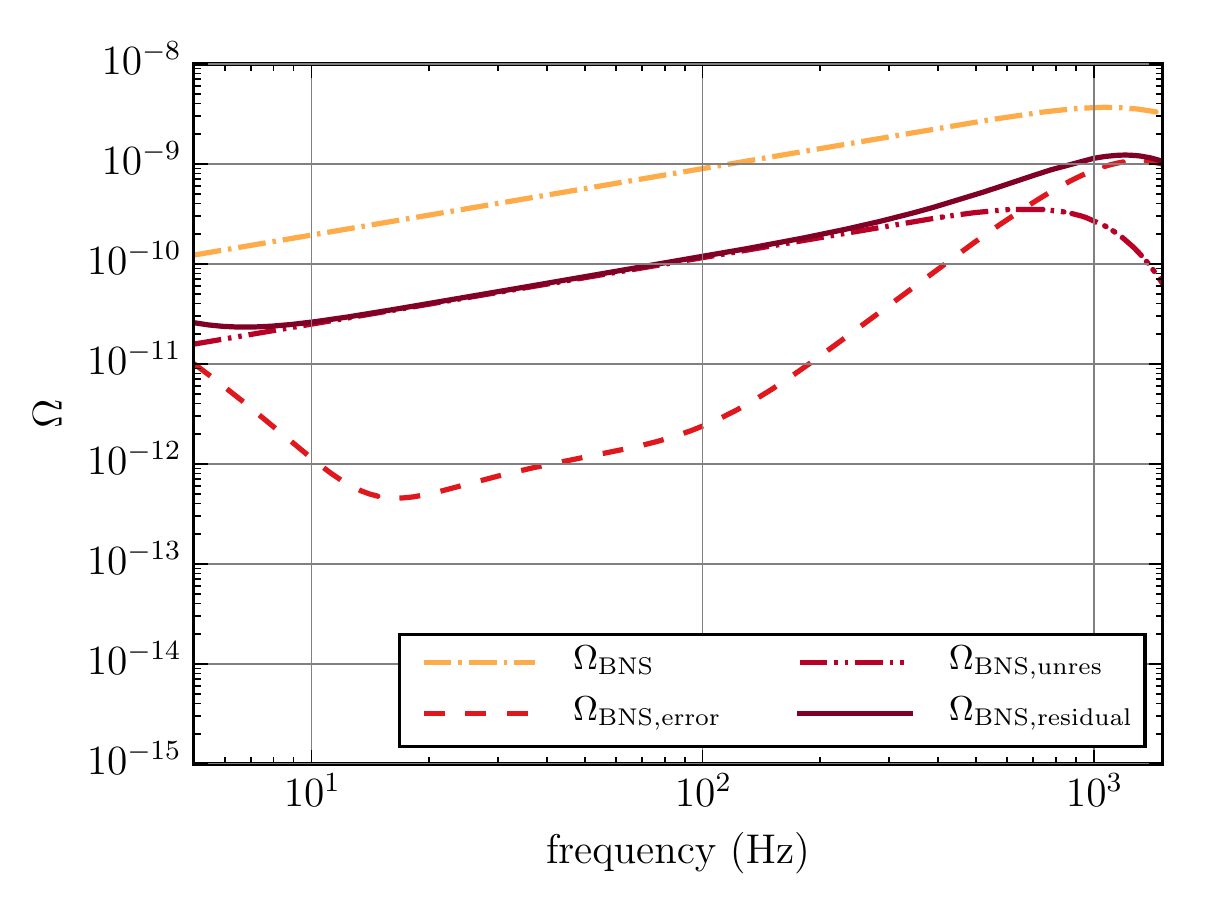} \caption{BNS, HLV network}
\end{subfigure}%
\hfill
\begin{subfigure}[t]{0.49\textwidth}
\includegraphics[width=1.0\textwidth]{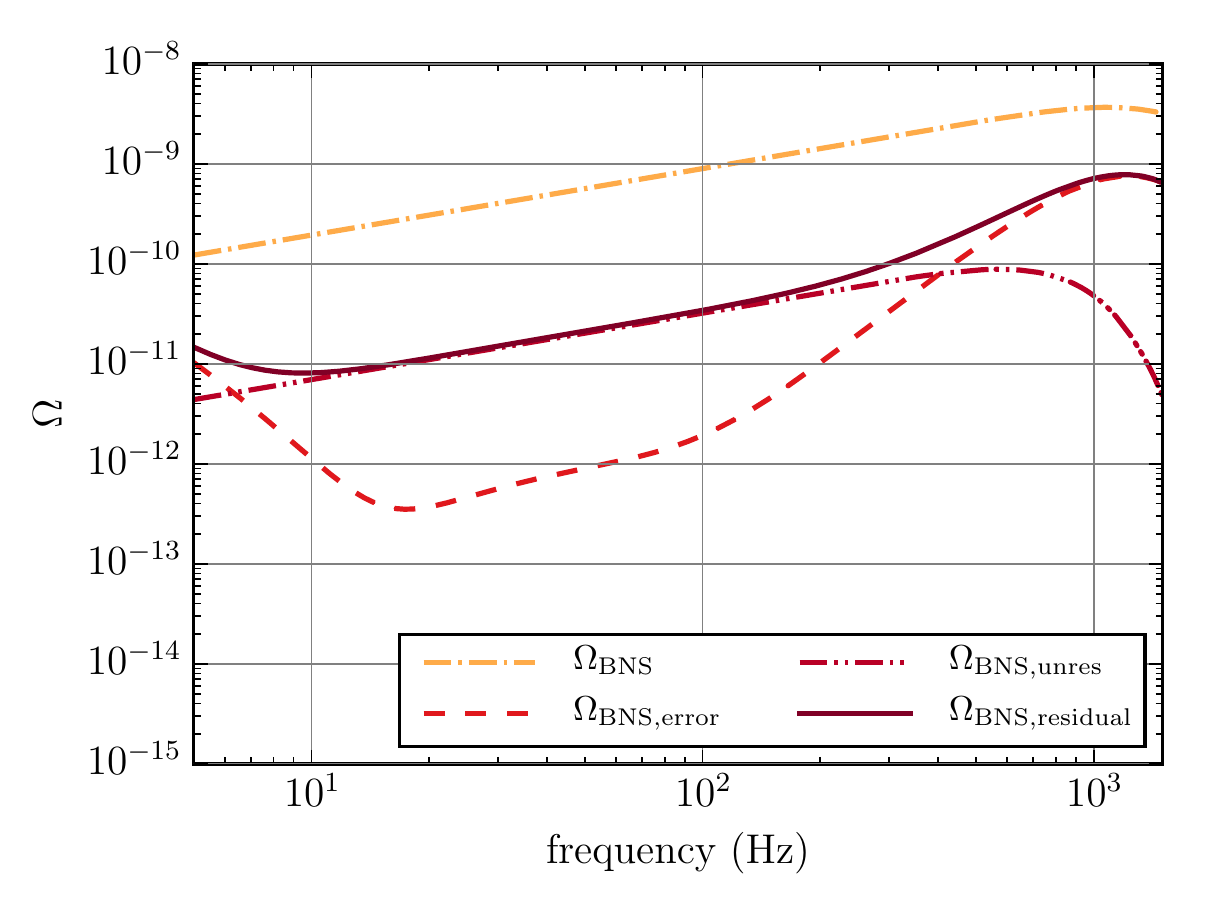} \caption{BNS, HLVIK network}
\end{subfigure}
\begin{subfigure}[t]{0.49\textwidth}
\includegraphics[width=1.0\textwidth]{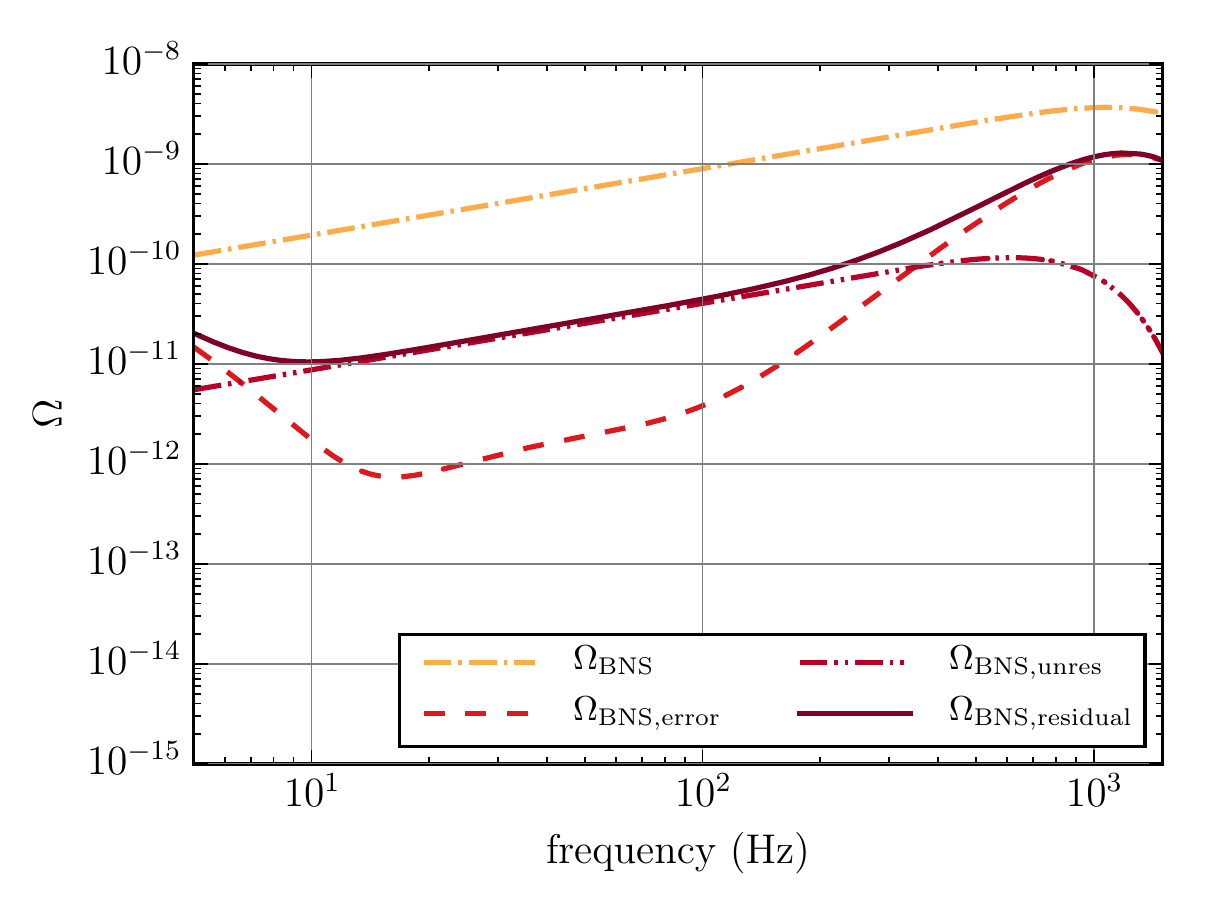}
\caption{BNS, HLV network, SNR threshold 8.0}
\end{subfigure}
\hfill
\begin{subfigure}[t]{0.49\textwidth}
\includegraphics[width=1.0\textwidth]{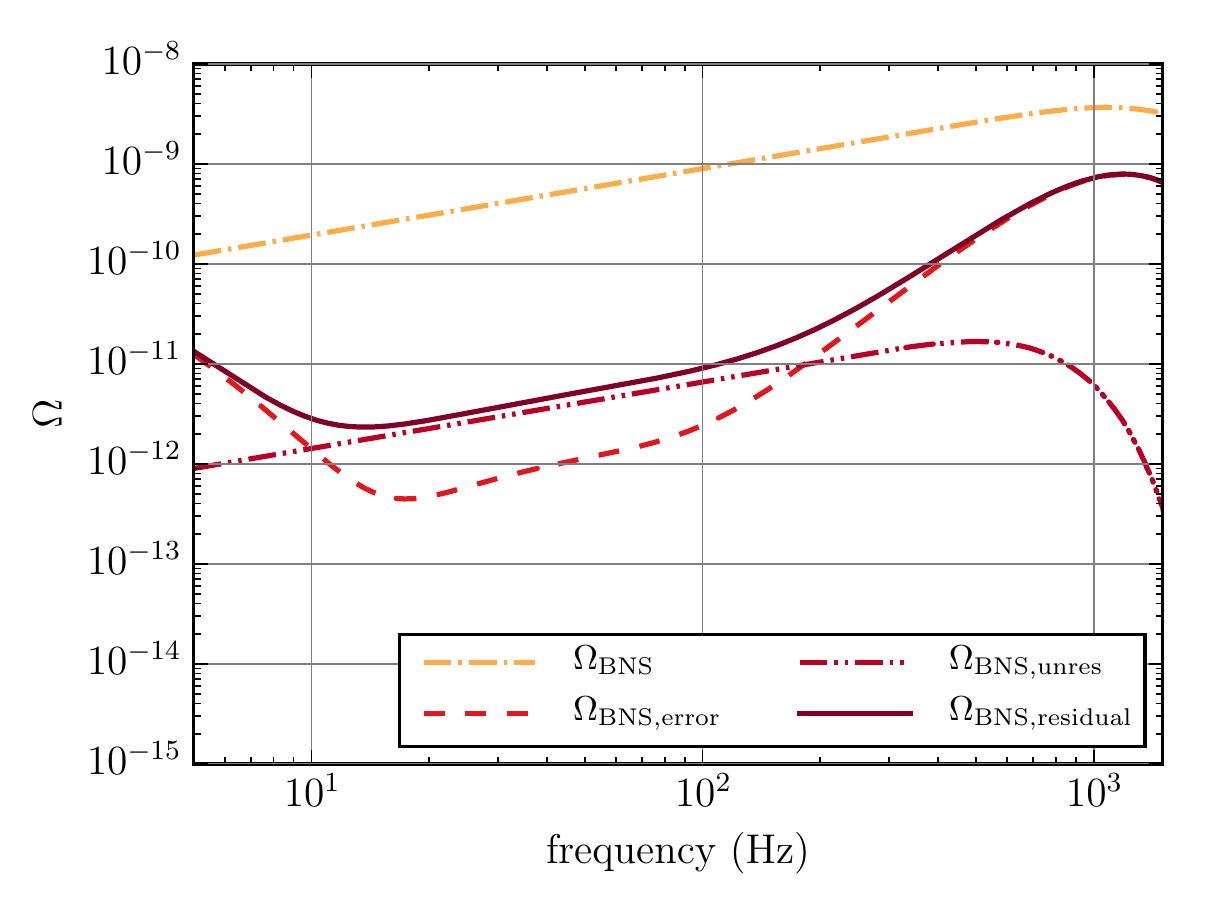}
\caption{BNS, HLVIK network, SNR threshold 8.0}
\end{subfigure}
\caption{The confusion background created by the astrophysical population of merging binary black holes (top two panels) and binary neutron stars (bottom four panels) is shown plotted (dot-dashed, orange lines) together with the background from unresolved sources (dot-dot-dashed, red lines), the background that remains after imperfect subtraction of resolved sources (dashed, red lines) and the sum of the latter two (solid, deep-red lines). The left panels are for a network of three 3G detectors and the right panels are for a network of five 3G detectors. We deem a source is resolved if the signal-to-noise it produces is $\ge 12$ for the top four panels, and $\ge 8$ for the bottom two panels.}
\label{fig:omega comparison}
\end{figure*}

\begin{figure}
\includegraphics[width=0.5\textwidth]{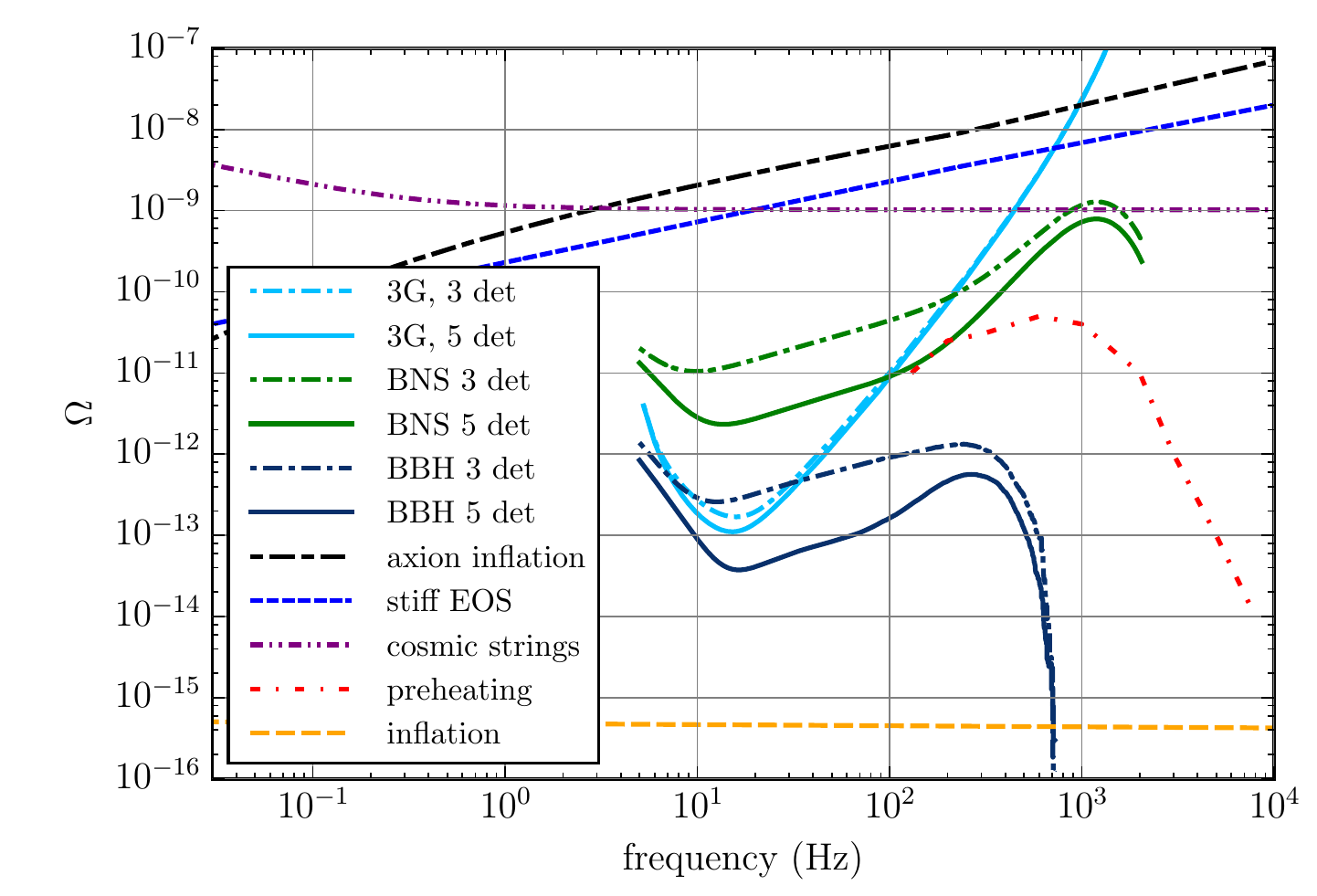}
\caption{Residual backgrounds after subtraction of the resolved foreground 
are plotted in solid (and dot-dashed) lines for a network of five 3G detectors 
(three 3G detectors, respectively) for the BNS cosmological population in green
and BBH population in deep blue. Also shown are the raw sensitivity curves for 
a stochastic background after one year of integration for a network of five
3G detectors (solid, cyan curves) and three 3G detectors (dot-dashed, cyan curves)
and the expected background from stiff equation-of-state, cosmic (super) strings,
preheating and inflation.}
\label{fig:Omega}
\end{figure}
\section{Simulations}\label{sec:simulations}
We simulate a population of BBH and BNS mergers according to the procedure
described in Sec.~\ref{sec:popsynth} for a year of data. There are 76,107 BBH and 1,438,835 BNS signals in our simulation. 
For each source, we calculate the expected network SNR assuming perfect template match, given by
\begin{equation}
\rho_i^\mathrm{net} = \sqrt{\sum_\mathrm{det}(\rho_i^\mathrm{det})^2},
\end{equation}
where index $i$ runs over all the sources, and $\rho_i^\mathrm{det}$
\begin{equation}
(\rho_i^\mathrm{det})^2 = 4 \int_0^\infty \mathrm{df} \frac{\abs{\tilde{h}_i^\mathrm{det}(f)}^2}{S_h^\mathrm{det}(f)}
\end{equation}
is the SNR for each source and detector pair ($i, \mathrm{det}$), and $\tilde{h}_i^\mathrm{det}(f) = F_+^\mathrm{det}\tilde{h}_{i, +} + F_\times^\mathrm{det}\tilde{h}_{i,\times}$ is the Fourier domain waveform projected on the detector.

We considered a source as resolvable and a part of the ``foreground'', whenever
$\rho_i^\mathrm{net}\geq \rho_\mathrm{thresh}=12.0$.  We use the 0 order PN
approximation for waveforms, since the results from that and a full
inspiral-merger-ringdown model agree to a great extent below 100 Hz.  It has
been shown for various detector combinations that frequencies below 100 Hz
account for more than 99\% of the SNR for the stochastic
search~\cite{2014PhRvD..89h4046R, prd.92.063002.15}.  Therefore for calculating
$\Omega_\mathrm{error}$, we only consider the 0th-PN model to compute the Fisher
matrix for each source in our simulation.

We calculate the Fisher matrices (and the variance-covariance matrices) for all
the sources in our simulation, and recover a set of parameters in order to
calculate $\Omega_\mathrm{residual, BNS}$ and $\Omega_\mathrm{residual, BBH}$.

Our results are shown plotted in Fig.\,\ref{fig:omega comparison}. For the
three-detector case, we find that 49\% of the BNS sources are unresolved (with
a network SNR $< 12$), whereas only 0.013\% of the BBH sources are unresolved.
For the five-detector case, we find that 25\% of the BNS sources are unresolved
while only 0.00075\% of the BBH sources remain unresolved. We show the results
for network SNR threshold of 12 in the first two rows of Fig.~\ref{fig:omega
comparison}. The first row shows the results for BBH (left: for a 3 detector 3G
network, right: for a 5 detector 3G network) and the second row shows the
results for the BNS (left: for a 3 detector 3G network, right: for a 5 detector
3G network). We can see that the $\Omega_\text{residual} =
\Omega_\text{error} + \Omega_\text{cbc, unres}$ depends on the network SNR
threshold. The higher the network SNR threshold, the lower the $\Omega_\text{error}$
but higher the $\Omega_\text{cbc, unres}$. Thus, the network SNR threshold can
be varied to minimize the $\Omega_\text{residual}$. 

For the BBH case, we have not tried to optimize the $\Omega_\text{residual, BBH}$, since 
it lies much below the $\Omega_\text{residual, BNS}$. For the BNS case, we can see from the
second row of Fig.~\ref{fig:omega comparison}, that we may be able to lower the
residual background by decreasing the network SNR threshold, since the residual
is dominated by the unresolved sources. We decided to lower the network SNR
threshold to $8$ (the threshold at which we should be able to resolve signals
in case of Gaussian noise); these results are shown in the last row of
Fig.~\ref{fig:omega comparison}. With a network SNR threshold of 8, the
number of unresolved BNS sources for a three (and five) network of 3G detectors
reduces to 25\% from 49\% (7.7\% from 25\%). We have managed to
lower the BNS residual background by lowering the detector network SNR threshold. The
residual background from the BNS sources still dominates over the BBH
background and is the limiting factor for the primordial backgrounds we can
observe.  An alternative would be to follow the noise projection method described in
Ref.~\cite{Harms:2008xv}, which does not require the SNR optimization procedure
described here.

\section{Discussion} \label{sec:discussion}
Conclusions of our study are summarized in Fig.\,\ref{fig:Omega}. The figure
plots the energy density in gravitational waves $\Omega(f)$ from axion
inflation \cite {Barnaby:2011qe}, a network of cosmic strings
\cite{prd.71.063510.05,prl.98.111101.07,prd.81.104028.10,prd.85.066001.12},
a background produced during post-inflation by oscillations of a fluid with an
equation-of-state stiffer than radiation \cite{Boyle:2007zx}, and from
post-inflation preheating scenarios \cite{Khlebnikov:1997di,Tilley:2000jh}
aided by parametric resonance \cite{prd.82.083518.10, Figueroa:2017vfa}.
For reference, we show the strength of the stochastic background from vacuum fluctuations
during standard inflation \cite{spjetp.40.409.75, jetpl.30.682.79, prd.48.3513.93}, although this will not be detectable by any of the
foreseen ground-based detector networks; others are examples of primordial backgrounds 
that could be potentially detected by 3G detectors. The strength of the background in these
examples depends on model parameters and it could be lower or higher than what is shown on the
plot.

The figure also shows the sensitivity of a network of three (and five) 3G
detectors to stochastic backgrounds assuming a one-year integration but in the
absence of confusion backgrounds from compact binaries or other astrophysical
populations. It is immediately apparent that the residual background, after
(imperfect) subtraction of the foreground sources, from binary neutron stars
will limit the strength of primordial backgrounds that could be detected by 3G
detectors. With a network of three (and five) 3G detectors, the sensitivity
will be limited to $\Omega_{\rm GW} \ge 10^{-11}$ at 10 Hz (respectively,
$\Omega_{\rm GW} \ge 3\times 10^{-12}$ at 15 Hz). The binary black hole
population, on the other hand, can be fully resolved and the residual from that
population has negligible effect on the raw sensitivity to stochastic
backgrounds. The rate of binary neutron stars could be larger or smaller than
the median rate of $R_m(z=0) = 920_{-790}^{+2220}
\textrm{Gpc}^{-3}\textrm{yr}^{-1}$ assumed in this paper, which would
correspondingly increase or decrease the confusion background of these sources.
Finally, increasing the number of 3G detectors from three to five improves the sensitivity 
to stochastic backgrounds by about factor of 5. This is accounted by the ability of the 
five-detector network to detect and subtract a greater number of sources; the volume
reach for a five-detector network increases by a factor $(5/3)^3 \sim 4.6$ relative 
to a three-detector network.

Keeping in mind that the strengths of the primordial backgrounds depend on the
specific model parameters that are not known, and the residual background could
vary based on the uncertainty in rate of compact binary mergers and the their mass
distribution, among other things, the figure shows the
most promising primordial background sources that this subtraction scheme could
reveal: cosmic strings, background from fluids with stiff EOS, and axion inflation.

\section*{Acknowledgements}
We thank Thomas Callister, Duncan Meacher and Alan Weinstein for helpful
discussions and comments. We thank Joe Romano for carefully reading the
manuscript and providing useful comments. We thank Andrew Matas for providing
useful data regarding some of the backgrounds considered in this paper. SS
acknowledges the support of the Eberly Research Funds of Penn State, The
Pennsylvania State University, University Park, PA. BSS was supported in part
by NSF grants PHY-1836779, AST-1716394 and AST-1708146 the Science and
Technology Facilities Council (STFC) of the United Kingdom. We acknowledge the
use of ICDS cluster at Penn State for the simulations in this work. This paper
has the LIGO document number LIGO-P2000009.

\bibliography{bibliography} 

\end{document}